\documentclass[13pt,a4paper]{article}
\usepackage{amsmath}
\usepackage{amsthm}
\usepackage{amsfonts}
\usepackage{amssymb}
\usepackage{euscript}
\usepackage[cp1251]{inputenc}
\usepackage[english] {babel}
\usepackage[pdftex]{graphicx}
\usepackage[12pt]{extsizes}

\begin{document}
\let\thefootnote\relax\footnote{This work was supported in part by the Simons Foundation.}
\begin{center}
  \textbf{AKNS hierarchy and Schrodinger finite-gap  potentials}
\end{center}
\begin{center}
  \textbf{V. Oganesyan}
\end{center}
\begin{center}
   \textbf{1. Introduction}
\end{center}

The commutativity condition $L_nL_m = L_mL_n$ of two differential operators
\begin{equation*}
L_n= \sum\limits^{n}_{i=0} u_i(x)\partial_x^i,  \quad  L_m= \sum\limits^{m}_{i=0} v_i(x)\partial_x^i,
\end{equation*}
where coefficients $u_i(x)$ and $v_i(x)$ are scalar or matrix valued functions, is equivalent to a very complicated system of nonlinear differential equations. The theory of commuting ordinary differential operators was first developed in the beginning of the XX century in the works of Wallenberg ~\cite{Wal}, Schur ~\cite{Schur}, and Burchnall, Chaundy ~\cite{Chaundy}. \\
If two differential operators with scalar or matrix valued coefficients commute, then there exists a nonzero polynomial $R(z,w)$ such that  $R(L_n,L_m)=0$ (see ~\cite{Chaundy}, ~\cite{Grin}). The curve $\Gamma$ defined by $R(z,w)=0$ is called the \emph{spectral curve} of this pair of operators.\\
If coefficients are scalar functions and
\begin{equation*}
L_n \psi=z\psi, \quad  L_m \psi=w\psi,
\end{equation*}
then $(z,w) \in \Gamma$. For almost all $(z,w) \in \Gamma$, the dimension of the space of common eigenfunctions $\psi$ is the same. The dimension of the space of common eigenfunctions of two commuting scalar differential operators is called the \emph{rank} of this pair. The rank is a common divisor of m and n. The genus of the spectral curve of a pair of commuting operators is sometimes called the genus
of this pair.\\
When coefficients are matrix of size $s\times s$ we have the \emph{vector rank} $(l_1,...,l_k)$, where $k\leqslant s$ (see ~\cite{Grin}). Numbers $l_i$ are common divisors of $m$ and $n$. \\
The first examples of pairs of commuting  scalar differential operators of the nontrivial ranks 2 and 3 and the nontrivial genus g=1 were constructed by Dixmier ~\cite{Dixmier} for the nonsingular elliptic spectral curve $w^2=z^3-\alpha$, where $\alpha$ is and arbitrary nonzero constant.\\
A general classification of commuting scalar  differential operators was obtained by Krichever ~\cite{ringkrichever}. Unfortunately this classification helps to find coefficients of commuting operators explicitly only when rank equals $1$ and in some other cases. The general form of commuting operators of rank 2 for an arbitrary elliptic spectral curve was found by Krichever and Novikov ~\cite{novkrich}. The general form of operators of rank 3 with arbitrary elliptic spectral curve was found by Mokhov ~\cite{Mokhov1},~\cite{Mokhov2}. Mironov in ~\cite{Mironov} found new methods of constructing new scalar commuting operators of rank 2. Using Mironov's method many examples of scalar commuting operators of rank 2 were found (see ~\cite{Mokhov4}, ~\cite{Vartan3} ~\cite{Zegl}, ~\cite{Vartan}, ~\cite{Vartan2}, ~\cite{Davl}).\\
A general classification of commuting matrix   differential operators  was obtained by Grinevich ~\cite{Grin}.\\
Many nonlinear equations are equivalent to commutativity condition of scalar or matrix differential operators. Using theory of commuting operators many exact solutions of nonlinear differential equations were found (see ~\cite{Smirnov},  ~\cite{Smirnov4}, ~\cite{Dubrovin1}, ~\cite{Mumford}, ~\cite{Dickey}).
\\
\\
The author is grateful to Professor O. I. Mokhov and Professor A. O. Smirnov for valuable discussions.
\begin{center}
\textbf{2. AKNS hierarchy and finite-gap Schrodinger potentials}
\end{center}

Let us consider
\begin{equation*}
M= \begin{pmatrix}
\imath & 0  \\
0 & -\imath  \\
\end{pmatrix}
\partial_x +
 \begin{pmatrix}
0 & -\imath q(x)  \\
\imath p(x) & 0  \\
\end{pmatrix}.
\end{equation*}
Denote by $L_n$ any differential operator of order $n$  with matrix coefficients of size $2\times 2$.\\
Define functions $f_k$, $g_k$ by the recurrence relations
\begin{equation}
 \begin{cases}
   f_{k+1}=\frac{i}{2}f'_{k}(x) - iq(x)(\int(p(x)f_{k}(x) + q(x)g_{k}(x))dx + C_k)\\
   g_{k+1}=-\frac{i}{2}g'_{k}(x) + ip(x)(\int(p(x)f_k(x) + q(x)g_k(x))dx + C_k)\\
 \end{cases}
\end{equation}
where $C_k$ are arbitrary constants and
\begin{equation*}
 \begin{cases}
   f_{1}=-iq(x)\\
   g_{1}=ip(x)\\
 \end{cases}
\end{equation*}
For example
\begin{equation*}
 \begin{cases}
   f_{3}= \frac{i}{4}q_{xx} - \frac{i}{2}pq^2 + \frac{C_1}{2}q_x - iC_2q\\
   g_{3}= - \frac{i}{4}p_{xx} + \frac{i}{2}p^2q + \frac{C_1}{2}p_x + iC_2p.\\
 \end{cases}
\end{equation*}
Easy to check that $f_k$ and $g_k$ are polynomials in $p,q,p',q',...p^{(k-1)},q^{(k-1)}$ (see \cite{weikard3}).  Functions $p(x), q(x)$ are solutions of
\begin{equation}
 \begin{cases}
   f_{m+1}=0\\
   g_{m+1}=0\\
 \end{cases}
\end{equation}
if and only if there exists an operator $L_{m+1}$ such that $ML_{m+1}=L_{m+1}M$ (see ~\cite{weikard3}) and
\begin{equation*}
L_{m+1}^2 = M^{2m+2} + b_{2m}M^{2m} + ... + b_1M + b_{0},
\end{equation*}
where $b_i$ are certain constants.
Easy to see from (1) that if condition (2) is satisfied, then
\begin{equation}
 \begin{cases}
   f_{k}=0\\
   g_{k}=0\\
 \end{cases}
\end{equation}
for all $k \geqslant m+1$.  \\
Assume that $p(x)$ and $q(x)$ have isolated poles of order $1$ at points $a_i$. And in a neighborhood of $a_i$ the Laurent series have the form
\begin{equation*}
p(x)=\frac{\varphi_{i,-1}}{x-a_i} + \varphi_{i,0} + \varphi_{i,1}(x-a_i)+...
\end{equation*}
\begin{equation*}
q(x)=\frac{\psi_{i,-1}}{x-a_i} + \psi_{i,0} + \psi_{i,1}(x-a_i)+...
\end{equation*}
The main results of this paper are the following.\\
\textbf{Theorem 1.}\\
\emph{If functions $p(x), q(x)$ are solutions of equations $f_{2m+2}=0$ and $g_{2m+2}=0$ or  $f_{2m+3}=0$ and $g_{2m+3}=0$, then}
\begin{equation*}
\begin{gathered}
\varphi_{i,-1} = \frac{n^2_i}{\psi_{i,-1}}, \varphi_{i,0} = -\frac{\psi_{i,0}\varphi_{i,-1}^2}{n^2_i},  \varphi_{i,1} = \frac{\psi_{i,1}\varphi_{i,-1}^2}{n^2_i}, \varphi_{i,2} = -\frac{\psi_{i,2}\varphi_{i,-1}^2}{n^2_i},\\ \varphi_{i,3} = \frac{\psi_{i,3}\varphi_{i,-1}^2}{n^2_i},..., \varphi_{i,2n_i-1} = \frac{\psi_{i,2n_i-1}\varphi_{i,-1}^2}{n^2_i},\\
\end{gathered}
\end{equation*}
\emph{where $n_i \in \mathbb{N}$ and $n_i \leqslant m$.}\\
\textbf{Theorem 2.}\\
\emph{If}
\begin{equation*}
\begin{gathered}
\varphi_{i,-1} = \frac{n^2_i}{\psi_{i,-1}}, \varphi_{i,0} = -\frac{\psi_{i,0}\varphi_{i,-1}^2}{n^2_i},  \varphi_{i,1} = \frac{\psi_{i,1}\varphi_{i,-1}^2}{n^2_i}, \varphi_{i,2} = -\frac{\psi_{i,2}\varphi_{i,-1}^2}{n^2_i},\\ \varphi_{i,3} = \frac{\psi_{i,3}\varphi_{i,-1}^2}{n^2_i},..., \varphi_{i,2n_i-1} = \frac{\psi_{i,2n_i-1}\varphi_{i,-1}^2}{n^2_i},\\
\end{gathered}
\end{equation*}
for some $n_i$, then solutions of
\begin{equation*}
\begin{pmatrix}
\imath & 0  \\
0 & -\imath  \\
\end{pmatrix}
\partial_xy +
 \begin{pmatrix}
0 & -\imath q(x)  \\
\imath p(x) & 0  \\
\end{pmatrix}y = zy
\end{equation*}
\emph{are meromorphic in a neighborhood of a pole $a_i$ for all  $z \in \mathbb{C}$, where $y=(y_1(x),y_2(x))^{t}$.}\\
\\
We know the following theorem (see ~\cite{weikard2})\\
\\
\textbf{Theorem.}\\
\emph{Assume that $p(x)$, $q(x)$ are rational functions bounded at infinity or else that $p(x)$, $q(x)$ are  meromorphic $\omega$-periodic functions with finitely many poles in the period strip. Then functions $p(x)$, $q(x)$ are solutions of some equations of AKNS hierarchy  if and only if all solutions of
\begin{equation*}
\begin{pmatrix}
\imath & 0  \\
0 & -\imath  \\
\end{pmatrix}
\partial_xy +
 \begin{pmatrix}
0 & -\imath q(x)  \\
\imath p(x) & 0  \\
\end{pmatrix}y = zy
\end{equation*}
are meromorphic for any  $z \in \mathbb{C}$, where $y=(y_1(x),y_2(x))^{t}$.}\\
\\
So, we obtain the following Theorem\\
\\
\textbf{Theorem 3.}\\
\emph{Assume that $p(x)$, $q(x)$ are rational functions bounded at infinity or else that $p(x)$, $q(x)$ are  meromorphic $\omega$-periodic functions with finitely many poles in the period strip. Assume that all poles of $p(x)$ and $q(x)$ are poles of order $1$. Then functions $p(x)$ and $q(x)$ are solutions of $f_{m}=0$ and $g_{m}=0$ if and only if}
\begin{equation*}
\begin{gathered}
\varphi_{i,-1} = \frac{n^2_i}{\psi_{i,-1}}, \varphi_{i,0} = -\frac{\psi_{i,0}\varphi_{i,-1}^2}{n^2_i},  \varphi_{i,1} = \frac{\psi_{i,1}\varphi_{i,-1}^2}{n^2_i}, \varphi_{i,2} = -\frac{\psi_{i,2}\varphi_{i,-1}^2}{n^2_i},\\ \varphi_{i,3} = \frac{\psi_{i,3}\varphi_{i,-1}^2}{n^2_i},..., \varphi_{i,2n_i-1} = \frac{\psi_{i,2n_i-1}\varphi_{i,-1}^2}{n^2_i}.\\
\end{gathered}
\end{equation*}
for some $n_i \in \mathbb{N}$.
\\
\\
Assume that $p(x)$, $q(x)$ satisfy the condition of Theorem 1. Let us consider the function $pq$. The Laurent series of function $pq$ in a neighborhood of $a_i$ has the form
\begin{equation*}
pq= \frac{h_{i,-2}}{(x-a_i)^2} + \frac{h_{i,-1}}{(x-a_i)} + h_{i,0} + h_{i,1}(x-a_i) + O((x-a_i)^2),
\end{equation*}
where
\begin{equation*}
\begin{gathered}
h_{i,-2} =n^2_i,\\
h_{i,2k-1} = (\varphi_{i,-1}\psi_{i,2k} + \psi_{i,-1}\varphi_{i, 2k}) + (\varphi_{i,0}\psi_{i,2k-1} + \psi_{i,0}\varphi_{i, 2k-1}) +...+\\
+(\varphi_{i,l}\psi_{i,2k-l-1} + \psi_{i,l}\varphi_{i, 2k-l-1}) + ...+(\varphi_{i,2k}\psi_{i,-1} + \psi_{i,2k}\varphi_{i, -1}).
\end{gathered}
\end{equation*}
But we know that $p$ and $q$ satisfy the condition of Theorem 1 and
\begin{equation*}
\varphi_{i,l}\psi_{i,2k-l-1} + \psi_{i,l}\varphi_{i, 2k-l-1} = \varphi_{i,-1}\psi_{i,-1}(\frac{\varphi_{i,l}\psi_{i,2k-l-1}}{\varphi_{i,-1}\psi_{i,-1}} + \frac{\psi_{i,l}\varphi_{i,2k-l-1}}{\varphi_{i,-1}\psi_{i,-1}} )=
\end{equation*}
\begin{equation*}
=\psi_{i,-1}(\frac{\psi_{i,2k-l-1}}{\psi_{i,-1}} + (-1)^{l+1}\frac{\varphi_{i,2k-l-1}}{\varphi_{i,-1}} ) = (\frac{\psi_{i,2k-l-1}}{\psi_{i,-1}} + (-1)^{2k-l-1}\frac{\varphi_{i,2k-l-1}}{\varphi_{i,-1}} ) = 0.
\end{equation*}
So, we see that coefficients $h_{i,-1} = h_{i,1}=...=h_{i,2n_i-1} =0$. We obtain that in a neighborhood of $a_i$ the Laurent series of function $pq$ have the form
\begin{equation*}
pq = \frac{n^2_i}{(x-a_i)^2} + \sum^{t=n_i}_{t=0}h_{i,2t}(x-a_i)^t +  h_{i, 2n_i+1}(x-a_i)^{2n_i+1} + O((x-a_i)^{2n_i+2})
\end{equation*}
\textbf{Theorem 4.} \emph{Let us suppose that $p(x)$ and $q(x)$ are rational or simply periodic functions with a common period. Let us assume that $p(x)$ and $q(x)$ haven't isolated singularity at infinity. Suppose that functions $p(x)$, $q(x)$ have finitely many poles in the period strip. Also suppose that all poles of $p,q$ are poles of order $1$ and $\varphi_{i,-1}\psi_{i,-1}=n^2$ for any $i$. If $p(x)$, $q(x)$ are solutions of $f_{m}=0$ and $g_{m}=0$ for any $m>2$, then function $u(x)=\frac{n+1}{n}pq$ is finite-gap Schrodinger  potential. }\\
\\
Let us consider some examples (see \cite{weikard3}, page 101)\\
\textbf{Example 1.}\\
Functions $p=\frac{\alpha}{\sin(x)}$ and $q=\frac{\beta}{\sin(x)}$, where $\alpha\beta=n^2$ and $n \in \mathbb{N}$, are solutions of $f_{2n+2}=0$ and $g_{2n+2}=0$. Product $pq=\frac{n^2}{\sin^2(x)}$  and  $\frac{n+1}{n}pq=\frac{n(n+1)}{\sin^2(x)}$ is Schrodinger finite-gap potential.\\
\textbf{Example 2.}\\
 Let us consider $p=\alpha(\zeta(x) - \zeta(x-\omega_2) - \zeta(\omega_2))$ and $q=\beta(\zeta(x) - \zeta(x-\omega_2) - \zeta(\omega_2))$, where $\zeta(x; \omega_1,\omega_2)$ is  Weierstrass function, $\alpha\beta = n^2$, $n \in \mathbb{N}$. Functions $p, q$ are solutions of  $f_{m+1}=0$ and $g_{m+1}=0$ for some $m$. Function $\frac{n+1}{n}pq=n(n+1)\wp(x) + n(n+1)\wp(x-\omega_2)$ is Schrodinger finite-gap potential.
\begin{center}
\textbf{3. Proof of Theorem 1}
\end{center}

We know that in a neighborhood of $a_i$
\begin{equation*}
\begin{gathered}
f_1=-\frac{i\psi_{i,-1}}{(x-a_i)} + O(1) = \frac{A^1_{i,-1}}{(x-a_i)} + O(1)\\
g_1=\frac{i\varphi_{i,-1}}{(x-a_i)} + O(1) = \frac{B^1_{i,-1}}{(x-a_i)} + O(1).
\end{gathered}
\end{equation*}
By $A^k_{i,m}$ denote a coefficient in the term $(x-a_i)^m$ in Laurent series  of $f_k$ at point $a_i$. Similarly by $B^k_{i,m}$ denote a coefficient in the term $(x-a_i)^m$ in Laurent series  of $g_k$ at point $a_i$.\\
\textbf{Lemma 1.} There exists $n_i\in \mathbb{N}$ such that $\varphi_{i,-1}=\frac{n^2_i}{\psi_{i,-1}}$ and $B^{k}_{i,-k} = (-1)^k\frac{\varphi_{i,-1}A^{k}_{i,-k}}{\psi_{i,-1}}$.\\
\textbf{Proof.}\\
We obviously have $B^1_{i,-1} = -\frac{\varphi_{i,-1}A^1_{i,-1}}{\psi_{i,-1}}$.\\
Let us calculate $f_2$ and $g_2$
\begin{equation*}
\begin{gathered}
f_2=-\frac{\psi_{i,-1}}{2(x-a_i)^2} + O(\frac{1}{(x-a_i)}) = \frac{A^2_{i,-2}}{(x-a_i)^2} + O(\frac{1}{(x-a_i)})\\
g_2=-\frac{\varphi_{i,-1}}{2(x-a_i)^2} + O(\frac{1}{(x-a_i)}) = \frac{B^2_{i,-2}}{(x-a_i)^2} + O(\frac{1}{(x-a_i)})
\end{gathered}
\end{equation*}
We get $B^2_{i,-2} = \frac{\varphi_{i,-1}A^2_{i,-2}}{\psi_{i,-1}}$. Consider $f_3$ and $g_3$
\begin{equation*}
\begin{gathered}
f_3=-\frac{i\psi_{i,-1}(\varphi_{i,-1}\psi_{i,-1}- 1)}{2(x-a_i)^3} + O(\frac{1}{(x-a_i)^2}) = \frac{A^3_{i,-3}}{(x-a_i)^3} + O(\frac{1}{(x-a_i)^2})\\
g_3=\frac{i\varphi_{i,-1}(\varphi_{i,-1}\psi_{i,-1} - 1)}{2(x-a_i)^3} + O(\frac{1}{(x-a_i)^2}) = \frac{B^3_{i,-3}}{(x-a_i)^3} + O(\frac{1}{(x-a_i)^2})
\end{gathered}
\end{equation*}
We get $B^3_{i,-3} = -\frac{\varphi_{i,-1}A^3_{i,-3}}{\psi_{i,-1}}$. Let us prove that
\begin{equation}
B^{2k+1}_{i,-2k-1} = -\frac{\varphi_{i,-1}A^{2k+1}_{i,-2k-1}}{\psi_{i,-1}}, \quad B^{2k+2}_{i,-2k-2} = \frac{\varphi_{i,-1}A^{2k+2}_{i,-2k-2}}{\psi_{i,-1}}.
\end{equation}
The proof is by induction on $k$. We checked this for $k=0$ and $k=1$. By the induction hypothesis,
\begin{equation*}
B^{2k+1}_{i,-2k-1} = -\frac{\varphi_{i,-1}A^{2k+1}_{i,-2k-1}}{\psi_{i,-1}}, \quad B^{2k}_{i,-2k} = \frac{\varphi_{i,-1}A^{2k}_{i,-2k}}{\psi_{i,-1}}.
\end{equation*}
We obtain
\begin{equation}
\begin{gathered}
f_{2k+2}=\frac{2i\psi_{i,-1}(A^{2k+1}_{i,-2k-1}\varphi_{i,-1} + B^{2k+1}_{i,-2k-1}\psi_{i,-1}) - iA^{2k+1}_{i,-2k-1}(2k+1)^2}{2(2k+1)(x-a_i)^{2k+2}} + O(\frac{1}{(x-a_i)^{2k+1}}) =\\
=-\frac{iA^{2k+1}_{i,-2k-1}(2k+1)}{2(x-a_i)^{2k+2}} + O(\frac{1}{(x-a_i)^{2k+1}}),
\end{gathered}
\end{equation}
\begin{equation}
\begin{gathered}
g_{2k+2}=\frac{iB^{2k+1}_{i,-2k-1}(2k+1)^2 - 2i\varphi_{i,-1}(A^{2k+1}_{i,-2k-1}\varphi_{i,-1} + B^{2k+1}_{i,-2k-1}\psi_{i,-1} ) }{2(2k+1)(x-a_i)^{2k+2}} + O(\frac{1}{(x-a_i)^{2k+1}}) =\\
=-\frac{iA^{2k+1}_{i,-2k-1}\varphi_{i,-1}(2k+1)}{2\psi_{i,-1}(x-a_i)^{2k+2}} + O(\frac{1}{(x-a_i)^{2k+1}})
\end{gathered}
\end{equation}
In this case  we see that $B^{2k+2}_{i,-2k-2} = \frac{\varphi_{i,-1}A^{2k+2}_{i,-2k-2}}{\psi_{i,-1}}$. Then
\begin{equation}
\begin{gathered}
f_{2k+3}=-i\frac{2(k+1)^2A^{2k+2}_{i,-2k-2} - \psi_{i,-1}(B^{2k+2}_{i,-2k-2}\psi_{i,-1} + A^{2k+2}_{i,-2k-2}\varphi_{i,-1})}{2(k+1)(x-a_i)^{2k+3}} + O(\frac{1}{(x-a_i)^{2k+2}}) =\\
=\frac{iA^{2k+2}_{i,-2k-2}(\varphi_{i,-1}\psi_{i,-1} - (k+1)^2)}{(k+1)(x-a_i)^{2k+3}} + O(\frac{1}{(x-a_i)^{2k+2}})
\end{gathered}
\end{equation}
\begin{equation}
\begin{gathered}
g_{2k+3}=i\frac{2(k+1)^2B^{2k+2}_{i,-2k-2} - \varphi_{i,-1}(B^{2k+2}_{i,-2k-2}\psi_{i,-1} + A^{2k+2}_{i,-2k-2}\varphi_{i,-1}) }{2(k+1)(x-a_i)^{2k+3}} + O(\frac{1}{(x-a_i)^{2k+2}}) =\\
=-\frac{iA^{2k+2}_{i,-2k-2}\varphi_{i,-1}(\varphi_{i,-1}\psi_{i,-1} - (k+1)^2)}{(k+1)\psi_{i,-1}(x-a_i)^{2k+3}} + O(\frac{1}{(x-a_i)^{2k+2}}).
\end{gathered}
\end{equation}
So (4) is proved.\\
We know that there exists $l$ such that $f_{l}=0$ and $g_{l}=0$ hence $A^l_{i,-l}=B^l_{i,-l}=0$ for some $l$. But we see from (5), (6) and (7), (8) that  $A^k_{i,-k}=B^k_{i,-k}=0$  if and only if $\varphi_{i,-1}\psi_{i,-1}=n^2_i$.\\
\textbf{The Lemma is proved.}\\
\\
\textbf{Lemma 2.} $\varphi_{i,0}=-\frac{\psi_{i,0}\varphi_{i,-1}^2}{n^2_i}$.\\
\textbf{Proof.}\\
As in the previous Lemma let us consider
\begin{equation*}
\begin{gathered}
f_1=-\frac{i\psi_{i,-1}}{(x-a_i)} -i\psi_{i,0} + O((x-a_i)) ,\\
g_1=\frac{i\varphi_{i,-1}}{(x-a_i)} + i\varphi_{i,0} + O((x-a_i)).
\end{gathered}
\end{equation*}
We see that $\frac{A^1_{i,0}}{\psi_{i,-1}} - \frac{B^1_{i,0}}{\varphi_{i,-1}} = -\frac{i}{n_i^2}(\psi_{i,-1}\varphi_{i,0} + \varphi_{i,-1}\psi_{i,0})$. Let us prove that
\begin{equation}
\frac{A^p_{i,-p+1}}{\psi_{i,-1}} + (-1)^p \frac{B^p_{i,-p+1}}{\varphi_{i,-1}} = \alpha_p(\psi_{i,-1}\varphi_{i,0} + \varphi_{i,-1}\psi_{i,0}),
\end{equation}
where $\alpha_p \neq 0$ and doesn't depend on $C_i$, $p\in \mathbb{N}$ is an arbitrary number. The proof is by induction on $p$. Suppose that the assumption is true for $A^p_{i,-p+1}$ and $B^p_{i,-p+1}$. Easy to see that
\begin{equation}
\begin{gathered}
A^{p+1}_{i,-p} = \frac{i\psi_{i,0}(B^p_{i,-p}\psi_{i,-1} + A^p_{i,-p}\varphi_{i,-1})}{p} +\\
+ \frac{i\psi_{i,-1}(A^p_{i,-p}\varphi_{i,0} + B^p_{i,-p}\psi_{i,0} + A^p_{i,-p+1}\varphi_{i,-1} + B^p_{i,-p+1}\psi_{i,-1})}{p-1} - \frac{iA^p_{i,-p+1}(p-1)}{2},
\end{gathered}
\end{equation}
\begin{equation}
\begin{gathered}
B^{p+1}_{i,-p} = -\frac{i\varphi_{i,0}(B^p_{i,-p}\psi_{i,-1} + A^p_{i,-p}\varphi_{i,-1})}{p} -\\
 -\frac{i\varphi_{i,-1}(A^p_{i,-p}\varphi_{i,0} + B^p_{i,-p}\psi_{i,0} + A^p_{i,-p+1}\varphi_{i,-1} + B^p_{i,-p+1}\psi_{i,-1})}{p-1} + \frac{iB^p_{i,-p+1}(p-1)}{2}.\\
\end{gathered}
\end{equation}
Suppose that $p$ is odd number and $p=2k+1$. By (4), Lemma 1 and the induction assumption  it follows that
\begin{equation}
\frac{A^{2k+2}_{i,-2k-1}}{\psi_{i,-1}} + \frac{B^{2k+2}_{i,-2k-1}}{\varphi_{i,-1}}  = -ik\alpha_{2k+1}(\varphi_{i,0}\psi_{i,-1} + \psi_{i,0}\varphi_{i,-1})
\end{equation}
If  $p$ is even and $p=2k$, then
\begin{equation*}
\begin{gathered}
\frac{A^{2k+1}_{i,-2k}}{\psi_{i,-1}} - \frac{B^{2k+1}_{i,-2k}}{\varphi_{i,-1}}  =\\
(\varphi_{i,0}\psi_{i,-1} + \psi_{i,0}\varphi_{i,-1})\frac{i((8k-2)A^{2k}_{i,-2k} + \alpha_{2k}k\psi_{i,-1}(4\varphi_{i,-1}\psi_{i,-1} -(2k-1)^2)}{2(2k-1)k\psi_{i,-1}}=\\
=(\varphi_{i,0}\psi_{i,-1} + \psi_{i,0}\varphi_{i,-1})\frac{i((8k-2)\frac{A^{2k}_{i,-2k}}{\psi_{i,-1}} + \alpha_{2k}k(4n_i^2 -(2k-1)^2)}{2(2k-1)k}=\\
=(\varphi_{i,0}\psi_{i,-1} + \psi_{i,0}\varphi_{i,-1})\frac{i((8k-2)\beta\prod\limits^{l=k-1}_{l=1}(n_i^2 - l^2) + \alpha_{2k}k(4n_i^2 -(2k-1)^2)}{2(2k-1)k},
\end{gathered}
\end{equation*}
where $\beta$ doesn't depend on $\varphi_{i,-1}$, $\psi_{i,-1}$ and $n_i$. Easy to see that  $\alpha_l \neq 0$ for any $n_i \in \mathbb{N}$ and $l$. We know that there exists $l$ such that $f_{l}=0$ and $g_{l}=0$. So $A^l_{i,-l+1} = B^l_{i,-l+1}=0 \Rightarrow \frac{A^l_{i,-l+1}}{\varphi_{i,-1}} + (-1)^l \frac{B^l_{i,-l+1}}{\psi_{i,-1}} = 0$. But we see that $\alpha_l \neq 0$ and hence $\frac{A^l_{i,-l+1}}{\varphi_{i,-1}} + (-1)^l \frac{B^l_{i,-l+1}}{\psi_{i,-1}} =0$ if and only if $\varphi_{i,0}\psi_{i,-1} + \psi_{i,0}\varphi_{i,-1} = 0$. \\
\textbf{The Lemma is proved.}\\
\\
Let us assume that
 \begin{equation}
 \psi_{i,-1}\varphi_{i,p} + (-1)^p\varphi_{i,-1}\psi_{i,p}=0, \quad \frac{A^{t}_{i,p-t+1}}{\psi_{i,-1}} + (-1)^{p-t}\frac{B^{t}_{i,p-t+1}}{\varphi_{i,-1}} =0 ,
 \end{equation}
where $p=0,...,k-1$ and $t\in \mathbb{N}$ is arbitrary number.\\
Obviously
\begin{equation*}
\frac{A^{1}_{i,k}}{\psi_{i,-1}} + (-1)^{k+1}\frac{B^1_{i,k}}{\varphi_{i,-1}} = -i(\frac{\psi_{i,k}}{\psi_{i,-1}} + (-1)^k\frac{\varphi_{i,k}}{\varphi_{i,-1}}) = -\frac{i}{n^2}(\varphi_{i,-1}\psi_{i,k} + (-1)^k\psi_{i,k}\varphi_{i,-1}).
\end{equation*}
Consider
\begin{equation*}
\begin{gathered}
f_r = \frac{A^r_{i,-r}}{(x-a_i)^r} + ... + A^r_{i,0} + A^r_{i,1}(x-a_i) + .. + A^r_{i,k-r}(x-a_i)^{k-r} + O((x-a_i)^{k-r+1})\\
g_r = \frac{B^r_{i,-r}}{(x-a_i)^r} + ... + B^r_{i,0} + B^r_{i,1}(x-a_i) + .. + B^r_{i,k-r}(x-a_i)^{k-r} + O((x-a_i)^{k-r+1})
\end{gathered}
\end{equation*}
and suppose that
\begin{equation}
\frac{A^{r-1}_{i,k-r+2}}{\psi_{i,-1}} + (-1)^{k-r+1}\frac{B^{r-1}_{i,k-r+2}}{\varphi_{i,-1}} =   \alpha_{r-1, k-r+2}(\varphi_{i,-1}\psi_{i,k} + (-1)^k\psi_{i,k}\varphi_{i,-1}),
\end{equation}
where $\alpha_{r-1, k-r+2}$ doesn't depend on $C_i$ and coefficients $\varphi_{i,j}$, $\psi_{i,j}$. But $\alpha_{r-1, k-r+2}$ depends on $n_i^2 = \varphi_{i,-1}\psi_{i,-1}$.\\
Let us prove that
\begin{equation}
\frac{A^{r}_{i,k-r+1}}{\psi_{i,-1}} + (-1)^{k-r}\frac{B^{r}_{i,k-r+1}}{\varphi_{i,-1}} = \alpha_{r, k-r+1}(\varphi_{i,-1}\psi_{i,k} + (-1)^k\psi_{i,k}\varphi_{i,-1}).
\end{equation}
Denote
\begin{equation*}
pf_{r-1} + qg_{r-1} = \frac{h_{i,-r}}{(x-a_i)^{r}} + ... + h_{i,k-r+1}(x-a_i)^{k-r+1} + O((x-a_i)^{k-r+1}).
\end{equation*}
Let us consider two cases.\\
1) $k - r$ is even number. \\
Using (13), we get \\
\begin{equation*}
\begin{gathered}
h_{i,k-r}=\varphi_{i,-1}A^{r-1}_{i,k-r+1} +\psi_{i,-1}B^{r-1}_{i,k-r+1} +...+\varphi_{i,k-1}A^{r-1}_{i,-r+1} + \psi_{i,k-1}B^{r-1}_{i,-r+1} =\\
=\varphi_{i,-1}\psi_{i,-1}((\frac{A^{r-1}_{i,k-r+1}}{\psi_{i,-1}} + \frac{B^{r-1}_{i,k-r+1}}{\varphi_{i,-1}})+...+(\frac{\varphi_{i,k-1}A^{r-1}_{i,-r+1}}{\varphi_{i,-1}\psi_{i,-1}} + \frac{\psi_{i,k-1}B^{r-1}_{i,-r+1}}{\psi_{i,-1}\varphi_{i,-1}}))=0
\end{gathered}
\end{equation*}
Similarly, under the condition of (13), we have $h_{i,2t}=0$, where $2t \leqslant k-r$.
From (1) it follows that
\begin{equation*}
\frac{f_{r}}{\psi_{i,-1}} + \frac{g_{r}}{\varphi_{i,-1}} = \frac{i}{2}(\frac{f'_{r-1}}{\psi_{i,-1}} - \frac{g'_{r-1}}{\varphi_{i,-1}}) +i(\frac{p}{\varphi_{i,-1}}-\frac{q}{\psi_{i,-1}})(\int(pf_{r-1} + qg_{r-1})dx + C_{r-1}).
\end{equation*}
Note that
\begin{equation}
\int(pf_{r-1} + qg_{r-1})dx = \sum\limits^{t=k-r}_{t = -r} \frac{h_{i,t}(x-a_i)^{t+1}}{t+1} +  \frac{h_{i,k-r+1}(x-a_i)^{k-r+1}}{k-r} + O((x-a_i)^{k-r+1}),
\end{equation}
where $h_{i,p}=0$, $p$ is even number and $p\leqslant k-r$. We mentioned before that $f_k$ and $g_k$ are polynomials in $p,q,p',q',...,p^{(k-1},q^{(k-1)}$ for all $k$. Hence $h_{i,-1}=0$.\\
Moreover, we see from (13) that
\begin{equation}
\frac{p}{\varphi_{i,-1}}-\frac{q}{\psi_{i,-1}} = \sum\limits_{t=0}^{2t < k}(\frac{\varphi_{i,2t}}{\varphi_{i,-1}} - \frac{\psi_{i,2t}}{\psi_{i,-1}} )(x-a_i)^{2t} + (\frac{\varphi_{i,k}}{\varphi_{i,-1}} - \frac{\psi_{i,k}}{\psi_{i,-1}})(x-a_i)^{k} + ...
\end{equation}
Combining (16) and (17) and using the fact that $k-r+1$ is odd number, we obtain
\begin{equation}
\begin{gathered}
\frac{A^{r}_{i, k-r+1}}{\psi_{i,-1}} + \frac{B^{r}_{i, k-r+1}}{\varphi_{i,-1}} = \frac{i(k-r+2)}{2}(\frac{A^{r-1}_{i,k-r+2}}{\psi_{i,-1}} - \frac{B^{r-1}_{i, k-r+2}}{\varphi_{i,-1}})= \\
= \alpha_{r-1, k-r+2}\frac{i(k-r+2)}{2}(\varphi_{i,-1}\psi_{i,k} + (-1)^k\psi_{i,-1}\varphi_{i,-1}).
\end{gathered}
\end{equation}
So, if $k-r$ is even, then assumption (15) is true.\\
\\
2) Suppose that $k-r$ is odd number.\\
Arguing as before we see that
\begin{equation*}
\begin{gathered}
h_{i,k-r+1}=\varphi_{i,-1}A^{r-1}_{i,k-r+2} +\psi_{i,-1}B^{r-1}_{i,k-r+2} +...+\varphi_{i,k-1}A^{r-1}_{i,-r+1} + \psi_{i,k-1}B^{r-1}_{i,-r+1} =\\
=\varphi_{i,-1}\psi_{i,-1}((\frac{A^{r-1}_{i,k-r+2}}{\psi_{i,-1}} + \frac{B^{r-1}_{i,k-r+2}}{\varphi_{i,-1}})+...+(\frac{\varphi_{i,k-1}A^{r-1}_{i,-r+2}}{\varphi_{i,-1}\psi_{i,-1}} + \frac{\psi_{i,k-1}B^{r-1}_{i,-r+2}}{\psi_{i,-1}\varphi_{i,-1}})+\\
+(\frac{\varphi_{i,k}A^{r-1}_{i,-r+1}}{\varphi_{i,-1}\psi_{i,-1}} + \frac{\psi_{i,k}B^{r-1}_{i,-r+1}}{\psi_{i,-1}\varphi_{i,-1}}))=\psi_{i,-1}B^{r-1}_{i,-r+1}(\frac{\varphi_{i,k}}{\varphi_{i,-1}} + (-1)^{r-1}\frac{\psi_{i,k}}{\psi_{i,-1}})=\\
=\beta\prod\limits^{l=[\frac{r-1}{2}]}_{l=1}(n^2 - l^2)(\frac{\varphi_{i,k}}{\varphi_{i,-1}} + (-1)^{k}\frac{\psi_{i,k}}{\psi_{i,-1}}),
\end{gathered}
\end{equation*}
where $\beta$ is constant and doesn't depend on coefficients of series.\\
Easy to see from (13) that $h_{i,2t+1}=0$, where $2t + 1 < k-r+1$. From (1) it follows that
\begin{equation*}
\frac{f_{r}}{\psi_{i,-1}} - \frac{g_{r}}{\varphi_{i,-1}} = \frac{i}{2}(\frac{f'_{r-1}}{\psi_{i,-1}} + \frac{g'_{r-1}}{\varphi_{i,-1}}) -i(\frac{p}{\varphi_{i,-1}}+\frac{q}{\psi_{i,-1}})(\int(pf_{r-1} + qg_{r-1})dx + C_{r-1}).
\end{equation*}
From (13) we obtain that
\begin{equation*}
\begin{gathered}
\frac{p}{\varphi_{i,-1}}+\frac{q}{\psi_{i,-1}} = \sum\limits_{t=-1}^{t=k-1 }(\frac{\varphi_{i,t}}{\varphi_{i,-1}} + \frac{\psi_{i,t}}{\psi_{i,-1}} )(x-a_i)^{t} + (\frac{\varphi_{i,k}}{\varphi_{i,-1}} + \frac{\psi_{i,k}}{\psi_{i,-1}})(x-a_i)^{k} + ...=\\
= \frac{2}{(x-a_i)} + 2\sum\limits_{t=0}^{2t \leqslant k-2}\frac{\varphi_{i,2t+1}}{\varphi_{i,-1}}(x-a_i)^{2t+1} + (\frac{\varphi_{i,k}}{\varphi_{i,-1}} + \frac{\psi_{i,k}}{\psi_{i,-1}})(x-a_i)^{k}...
\end{gathered}
\end{equation*}
Using the fact that $k-r+1$ is even number, we get
\begin{equation*}
\begin{gathered}
\frac{A^{r}_{i, k-r+1}}{\psi_{i,-1}} - \frac{B^{r}_{i, k-r+1}}{\varphi_{i,-1}} = \frac{i(k-r+2)}{2}(\frac{A^{r-1}_{i,k-r+2}}{\psi_{i,-1}} + \frac{B^{r-1}_{i, k-r+2}}{\varphi_{i,-1}}) -2\frac{h_{i,k-r+1}}{k-r+1}=\\
= (\frac{i(k-r+2)\alpha_{r-1, k-r+2}}{2} - 2\beta\prod\limits^{l=[\frac{r-1}{2}]}_{l=1}(n^2 - l^2) )(\frac{\varphi_{i,k}}{\psi_{i,-1}} + (-1)^{k}\frac{\psi_{i,k}}{\varphi_{i,-1}}).
\end{gathered}
\end{equation*}
We obtain that if $k-r$ is odd, then assumption (15) is correct.\\
So, we proved that assumption (15) is correct for all $k$ and $r$.\\
\\
Suppose that $k<2n_i+1$. We know from Lemma 1 that $f_k \neq 0$ and $g_k\neq 0$ because  $A^k_{i,-k} \neq 0$.
We obtain that if there exists $m$ such that $f_{2m+1}=0$ and $g_{2m+1}=0$, then $A^{2m+1}_{i,p}=B^{2m+1}_{i,p}=0 \Rightarrow A^{2m+1}_{i,p} + (-1)^{p+1}B^{2m+1}_{i,p}=0$, where $p \leqslant 0$, $m\geqslant n_i$. Note that $A^{2n_i+1}_{i,-2n_i-1}=B^{2n_i+1}_{i,2n_i+1}=0$. \\
\\
We see from (18) that $\frac{A^{2n_i+2}_{i,-1}}{\psi_{i,-1}} + \frac{B^{2n_i+2}_{i,-1}}{\varphi_{i,-1}}=0$. We get from (15) that
\begin{equation*}
\frac{A^{2m+1}_{i,k-2m}}{\psi_{i,-1}} + (-1)^{k-2m+1}\frac{B^{2m+1}_{i,k-2m}}{\varphi_{i,-1}} =   \alpha_{2m+1, k-2m }(\frac{\varphi_{i,k}}{\psi_{i,-1}} + (-1)^{k}\frac{\psi_{i,k}}{\varphi_{i,-1}})=0.
\end{equation*}
Easy to see that $\alpha_{2m+1, k-2m}\neq 0$, where $k-2m<-1$. This means that $\frac{\varphi_{i,k}}{\psi_{i,-1}} + (-1)^{k}\frac{\psi_{i,k}}{\varphi_{i,-1}} = 0 \Rightarrow \varphi_{i,k}=(-1)^{k+1}\frac{\psi_{i,k}\varphi_{i,-1}^2}{n^2}$ for any $k \leqslant 2n_i-1$.\\
\\
\textbf{Theorem 1 is proved.}
\\
\begin{center}
\textbf{4. Proof of Theorem 2}
\end{center}

Let us consider the equation
\begin{equation}
M= \begin{pmatrix}
\imath & 0  \\
0 & -\imath  \\
\end{pmatrix}
\partial_xy +
 \begin{pmatrix}
0 & -\imath q(x)  \\
\imath p(x) & 0  \\
\end{pmatrix}y = zy
\end{equation}
where $y=(y_1(x),y_2(x))^{t}$. Solution of (19) have poles only at poles of $p$ and $q$. \\
Assume that
\begin{equation*}
\begin{gathered}
y_1=\alpha_0(x-a_i)^{\sigma} + \alpha_1(x-a_i)^{\sigma+1}+...\\
y_2=\beta_0(x-a_i)^{\sigma} + \beta_1(x-a_i)^{\sigma+1}+...
\end{gathered}
\end{equation*}
where $a_i$ is a pole of functions $p(x)$ and $q(x)$. Substituting $y$ in (19) and considering the coefficient of $(x-a_i)^{\sigma-1}$ we get
\begin{equation*}
 \begin{cases}
   \sigma\alpha_0- \psi_{i,-1}\beta_0=0\\
   \varphi_{i,-1}\alpha_0 -\sigma\beta_0  =0.\\
 \end{cases}
\end{equation*}
There exist solutions $\alpha_0, \beta_0 \neq 0$ if and only if $\sigma = \pm n_i$, where $n_i \in \mathbb{N}$.  Consider the coefficient of $(x-a_i)^{k+\sigma}$,  where $k$ is integer and $k\geqslant 0$. We obtain
\begin{equation*}
 \begin{cases}
   (k+\sigma)\alpha_{k-n_i+1}- \psi_{i,-1}\beta_{k-n_i+1}=F_1\\
   \varphi_{i,-1}\alpha_{k-n_i+1} -(k+\sigma)\beta_0 =F_2\\
 \end{cases}
\end{equation*}
where $F_1$ and $F_2$ don't depend  on $\alpha_{k-n_i+1}$ and $\beta_{k-n_i+1}$. Determinant of this system is equal to zero if and only if $k = -2\sigma$ and $k=0$. So, we have problems with finding coefficients only when $\sigma=-n_i$ and $k=2n_i$ because $k\geqslant 0$.\\
Let us prove that if $k\leqslant 2n_i-1$, then
\begin{equation}
n_i\beta_k + (-1)^k\varphi_{i,-1}\alpha_k =0
\end{equation}
or equivalently
\begin{equation*}
n_i\alpha_k + (-1)^k\psi_{i,-1}\beta_k=0.
\end{equation*}
We see that  $n_i\alpha_0 + \psi_{i,-1}\beta_0=0$ or equivalently $\varphi_{i,-1}\alpha_0 + n\beta_0=0$. Then
\begin{equation}
\begin{cases}
  (1-n_i)\alpha_1 - \psi_{i,-1}\beta_1  - \psi_{i,0}\beta_0=z\alpha_0\\
   (n_i-1)\beta_1 + \varphi_{i,-1}\alpha_1 + \varphi_{i,0}\alpha_0 =z\beta_0.\\
 \end{cases}
\end{equation}
Multiplying the first equation of (21) by $n_i$, the second by $\psi_{i,-1}$ and summing, we get
\begin{equation*}
n_i\alpha_{1} - \psi_{i,-1}\beta_1 + \psi_{i,-1}\varphi_{i,0}\alpha_0 - n\psi_{i,0}\beta_0=0
\end{equation*}
Under the condition of the Theorem we obtain that $n_i\alpha_{1} - \psi_{i,-1}\beta_1=0$. So, if $k=1$, then assumption (20) is true. Similarly, let us prove (20) for $k+1$ assuming that (20) is true for $k$, where $k< 2n_i-1$. Calculations show\\
\begin{equation}
\begin{cases}
  (n_i - k)\alpha_k - \psi_{i,-1}\beta_k - \psi_{i,0}\beta_{k-1}-...- \psi_{i,k-1}\beta_0=z\alpha_{k-1}\\
   -(n_i - k)\beta_k + \varphi_{i,-1}\alpha_k + \varphi_{i,0}\alpha_{k-1} + ...+\varphi_{i,k-1}\alpha_0 =z\beta_{k-1}.\\
 \end{cases}
\end{equation}
Let us multiply the first equation of (22) by $n_i$ the second by $(-1)^{k+1}\psi_{i,-1}$  and add. Easy to see that under the conditions of the Theorem we get (20).\\
Now let us consider coefficient of $(x-a_i)^{n_i-1}$
\begin{equation*}
\begin{cases}
n_i\alpha_{2n_i} - \psi_{i,-1}\beta_{2n_i} - \psi_{i,0}\beta_{2n_i-1} - ... -\psi_{2n_i-1}\beta_0 = z\alpha_{2n_i-1}\\
 \varphi_{i,-1}\alpha_{2n_i} + \varphi_{i,0}\alpha_{2n_i-1} + ... +\varphi_{2n_i-1}\alpha_0 - n\beta_{2n_i} = z\beta_{2n_i-1}\\
 \end{cases}
\end{equation*}
\begin{equation*}
\begin{cases}
n_i\alpha_{2n_i}  = z\alpha_{2n_i-1} + \psi_{i,-1}\beta_{2n_i} +  \psi_{i,0}\beta_{2n_i-1} + ... +\psi_{2n_i-1}\beta_0\\
  \varphi_{i,-1}\alpha_{2n_i} = n_i\beta_{2n_i} -  \varphi_{i,0}\alpha_{2n_i-1} - ... -\varphi_{2n_i-1}\alpha_0 + z\beta_{2n_i-1}.  \\
 \end{cases}
\end{equation*}
This system has solutions if and only if
\begin{equation*}
\begin{gathered}
z \frac{\alpha_{2n_i-1}}{n_i} + \frac{\psi_{i,-1}\beta_{2n_i}}{n_i} +  \frac{\psi_{i,0}\beta_{2n_i-1}}{n_i} + ... +\frac{\psi_{2n_i-1}\beta_0}{n_i} -\\ -n_i\frac{\beta_{2n_i}}{\varphi_{i,-1}} -  \frac{\varphi_{i,0}\alpha_{2n_i-1}}{\varphi_{i,-1}} - ... -\frac{\varphi_{2n_i-1}\alpha_0}{\varphi_{i,-1}} - z\frac{\beta_{2n_i-1}}{\varphi_{i,-1}}=0.\\
\end{gathered}
\end{equation*}
But we know from Lemma 1 that $\varphi_{i,-1}\psi_{i,-1}=n^2$ and under the conditions of the Theorem and (20) we have
\begin{equation*}
\begin{gathered}
\beta_{2n_i}(\frac{\varphi_{i,-1}\psi_{i,-1} -n^2_i}{n_i\varphi_{i,-1}}) + \frac{\varphi_{i,-1}\psi_{i,0}\beta_{2n_i-1} - n_i\varphi_{i,0}\alpha_{2n_i-1}}{n_i\varphi_{i,-1}} +...+ \frac{\varphi_{i,-1}\psi_{i,2n_i-1}\beta_{0} - n_i\varphi_{i,2n_i-1}\alpha_{0}}{n_i\varphi_{i,-1}} +\\ +z(\frac{\varphi_{i,-1}\alpha_{2n_i-1} - n_i\beta_{2n_i-1}}{n_i\varphi_{i,-1}}) = \\
=\varphi_{i,0}(\frac{\psi_{i,-1}\beta_{2n_i-1} - n_i\alpha_{2n_i-1}}{n_i\varphi_{i,-1}}) +...+ \varphi_{i,2n_i-1}\frac{-\psi_{i,-1}\beta_{0} -n_i\alpha_{0}}{n_i\varphi_{i,-1}} +\\ +z(\frac{\varphi_{i,-1}\alpha_{2n_i-1} - n_i\beta_{2n_i-1}}{n_i\varphi_{i,-1}})=0.
\end{gathered}
\end{equation*}
So, $y=(y_1(x),y_2(x))^{t}$, where $\sigma=\pm n_i$, are two linear independent formal solutions then it is also an actual solution near $a_i$ (see \cite{Levinson}, \S 4.3).
Finally, we obtain that solutions of (21) are locally meromorphic for all $z \in \mathbb{C}$.
\\
\textbf{Theorem is proved.}
\\
\\
\begin{center}
\textbf{5. Proof of Theorem 4}
\end{center}

Suppose $u(x)$ is a rational function  or a simply periodic function or an elliptic function. Suppose that $u(x)$ hasn't isolated singularity at infinity. Assume that $u(x)$  has finitely many poles in the  period strip or fundamental parallelogram. Suppose that function $u(x)$ has poles at points $a_1,a_2,..a_{N}$. In a neighborhood of $a_i$
\begin{equation*}
u(x) = \frac{a_{i,-2}}{(x-a_i)^2} + \frac{a_{i,-1}}{(x-a_i)} + a_{i,0} + a_{i,1}(x-a_i) + O((x-a_i)^2).
\end{equation*}
Theorem 3 follows from the following theorem ~\cite{weikard1}\\
\textbf{Theorem.}\\
\emph{Function $u(x)$ is Schrodinger  finite-gap potential if and only if
\begin{equation*}
a_{i,-2}=n_i(n_i+1), a_{i,-1}=0, a_{i,1}=0, ... , a_{2n_i - 1} = 0.
\end{equation*}}
But we have already proved that in a neighborhood of poles $a_i$ the Laurent series of function $\frac{n+1}{n}pq$  have the form
\begin{equation*}
\frac{n_i(n_i+1)}{(x-a_i)^2} + \sum^{t=n_i}_{t=0}h_{i,2t}(x-a_i)^{2t} + h_{2n_i+1}(x-a_i)^{2n_i+1} + O((x-a_i)^{2n_i+2})
\end{equation*}
\textbf{Theorem 3 is proved.}\\

Department of Geometry and Topology, Faculty of Mechanics and Mathematics, Lomonosov Moscow State University, Moscow, 119991 Russia.\\\\
E-mail address: vardan.o@mail.ru

\end{document}